\documentclass[a4paper]{article}

\usepackage{INTERSPEECH2022}
\usepackage{multirow}
\usepackage{tabularx}
\usepackage{booktabs}
\usepackage{cite}
\usepackage{subcaption}

\title{Learnable Frequency Filters for Speech Feature Extraction in Speaker Verification}
\name{Jingyu Li, Yusheng Tian, Tan Lee}
\address{Department of Electronic Engineering, The Chinese University of Hong Kong, Hong Kong}
\email{\{lijingyu0125, ystian0617\}@link.cuhk.edu.hk, tanlee@ee.cuhk.edu.hk}

\begin{document}

\maketitle
\begin{abstract}
Mel-scale spectrum features are used in various recognition and classification tasks on speech signals. There is no reason to expect that these features are optimal for all different tasks, including speaker verification (SV). This paper describes a learnable front-end feature extraction model. The model comprises a group of filters to transform the Fourier spectrum. Model parameters that define these filters are trained end-to-end and optimized specifically for the task of speaker verification. Compared to the standard Mel-scale filter-bank, the filters’ bandwidths and center frequencies are adjustable. Experimental results show that applying the learnable acoustic front-end improves speaker verification performance over conventional Mel-scale spectrum features. Analysis on the learned filter parameters suggests that narrow-band information benefits the SV system performance. The proposed model achieves a good balance between performance and computation cost. In resource-constrained computation settings, the model significantly outperforms CNN-based learnable front-ends. The generalization ability of the proposed model is also demonstrated on different embedding extraction models and datasets.
\end{abstract}

\section{Introduction}
\label{sec:intro}
Speaker verification (SV) refers to the task of verifying the identity of a speaker from given speech utterance(s). SV systems are developed for various notable applications\cite{reynolds2002overview,singh2012applications,kim2019deep}, such as speaker diarization, bio-metric authentication, and security. Deep neural network (DNN) based models \cite{snyder2017deep,snyder2018x,nagrani2020voxceleb,chung20b_interspeech,desplanques20_interspeech} are predominantly adopted in current SV systems and lead to appreciable performance gain over
conventional models, e.g., GMM-UBM, I-vectors\cite{reynolds2000speaker,zheng2004text,dehak2010front}.
Typically these DNN models take a certain form of acoustic features as input and produce neural embeddings that represent speaker-specific information in speech, which are then used for speaker discrimination. The most commonly used input acoustic features are Mel-scale spectrum features like log Mel-scale filter-bank coefficients (\emph{MFBANK}) and Mel-Frequency Cepstral Coefficients (\emph{MFCC}). They are computed from Short-Time Fourier Transform (STFT) coefficients and transformed using a set of pre-defined band-pass filters designed with consideration on human auditory perception \cite{stevens1940relation}. These acoustic features are widely used and have achieved great success across different tasks of speech and language processing. However, there is no reason to expect that this universal acoustic front-end is optimal and performs equally well for a specific task like speaker verification. In \cite{ravanelli2018speaker}, it is argued that narrow-band spectral information may contain distinct characteristics of speakers, and Mel-scale spectrum features might have ignored lots of narrow-band information.

Could we improve the performance of an SV system by learning the audio front-end as part of model training?  In other application areas of deep learning, e.g., computer vision (CV), it has been shown that feature representations learned from raw input, i.e., image pixels, perform better than hand-crafted features in various modeling and classification tasks \cite{krizhevsky2012imagenet,DBLP:journals/corr/SimonyanZ14a,szegedy2015going}. There were also a number of studies by the speech research community on applying CNN to learn features from raw waveform in conjunction with the downstream task \cite{hoshen2015speech,sainath15_interspeech,muckenhirn2018towards}. Experimental results show their superior performance to hand-crafted features like \emph{MFBANK} and \emph{MFCC}. However, the performance gain is at the cost of significantly increased computation that is due to the small-stride CNN.

In this paper, a computationally efficient learnable acoustic front-end is proposed for SV systems. The front-end consists of a group of learnable filters that extract features with low computation cost by directly transforming the STFT spectrum. These \textbf{Learnable Frequency-Filters} (\emph{LFF}) are similar to Mel-scale filters but allow flexible adjustment on the filters' frequency responses.
The filters' bandwidths and center frequencies are updated in conjunction with the embedding extraction model in an end-to-end manner. Experimental results show that the proposed method achieves better performance than \emph{MFBANK} and two learnable CNN-based feature extraction models. By analyzing the learned filters, it is noted that the flexibly adjusted bandwidth accounts for most of the improvement, while the learned center frequencies are very similar to those used in Mel-scale filterbank.

The remainder of the paper is organized as follows. The relation to previous works is described in Section~\ref{sec:related}. Section 3 discusses the architecture of the proposed model. Experimental setup and results are given in Sections 4 and 5. Finally, Section 6 contains discussions and conclusions.

\section{Previous works}
\label{sec:related}
The relation between \emph{MFBANK} and the convolution layer is discussed first, in order to relate conventional signal analysis to feature learning in CNN. Then we give a brief review on different learnable feature approaches in previous research.

\subsection{Mel-scale filter-bank features}
\label{ssec:MelAF}
The computation of \emph{MFBANK} consists of two major parts: $(1)$ STFT and $(2)$ Mel-scale filter-banks. The STFT spectrum of input sample $X$ is composed of a sequence of Fourier transform (FT) coefficients, which are calculated on the short-time frame sequence $[x_1,...x_n]$. These frames are cropped from $X$ with window length $w$ and hop length $s$. A window function $f_{window}$, e.g., Hanning window, is applied to each frame. The FT coefficient on frequency bin $k$ can be represented by the dot product between time-domain signal samples and a complex-valued filter $f_k=e^{-2\pi i kt/N}$, where $N$ is the number of frequency bins in the FT and $t$ is the time index in a frame. Thus, the STFT coefficients on the frame $x_j$ can be expressed as,
\begin{equation}
  y[j, k] = \left <x_j \odot f_{window},  f_k\right>, j \in [1, n],k \in [0, N/2]
  \label{eq:ft}
\end{equation}
where $\odot$ and $\left<,\right>$ denote the operation of element-wise product and inner product respectively. And it can be written in the convolution form as:
\begin{equation}
  Y[k] = X * (f_k\odot f_{window})
  \label{eq:stft}
\end{equation}
where $*$ is convolution operation with stride $s$. In this way STFT is made equivalent to a convolution layer in DNN with kernel size $w$, stride $s$, and output channel $N/2$, with fixed complex-valued weights. A common setting of STFT at sampling rate of $16kHz$ audio uses $25ms$ window length and $10ms$ hop length, which correspond to $w$=$400$ and $s$=$160$. The Mel-scale filters are applied on the STFT spectrum for the \emph{MFBANK} features.

\subsection{Learnable feature front-end}
\label{ssec:Lfeat}

\subsubsection{Vanilla CNN filters}
\label{sssec:LFCNN}
Given that STFT and the filterbank operation can be represented as convolution, a number of studies proposed to use learnable convolution kernels to generate filter-banks\cite{sainath15_interspeech,zeghidour2018learning,zeghidour18_interspeech}. They showed that the filter-bank structure learned from CNN can approximate Mel-scale filter-bank with properly initialized weights, suggesting that learnable filter-banks have the ability to outperform \emph{MFBANK}, or at least get close to it.

\subsubsection{Parameterized CNN filters}
\label{sssec:LFPower}
\begin{figure}[t]
  \centering
  \includegraphics[width=0.85\linewidth]{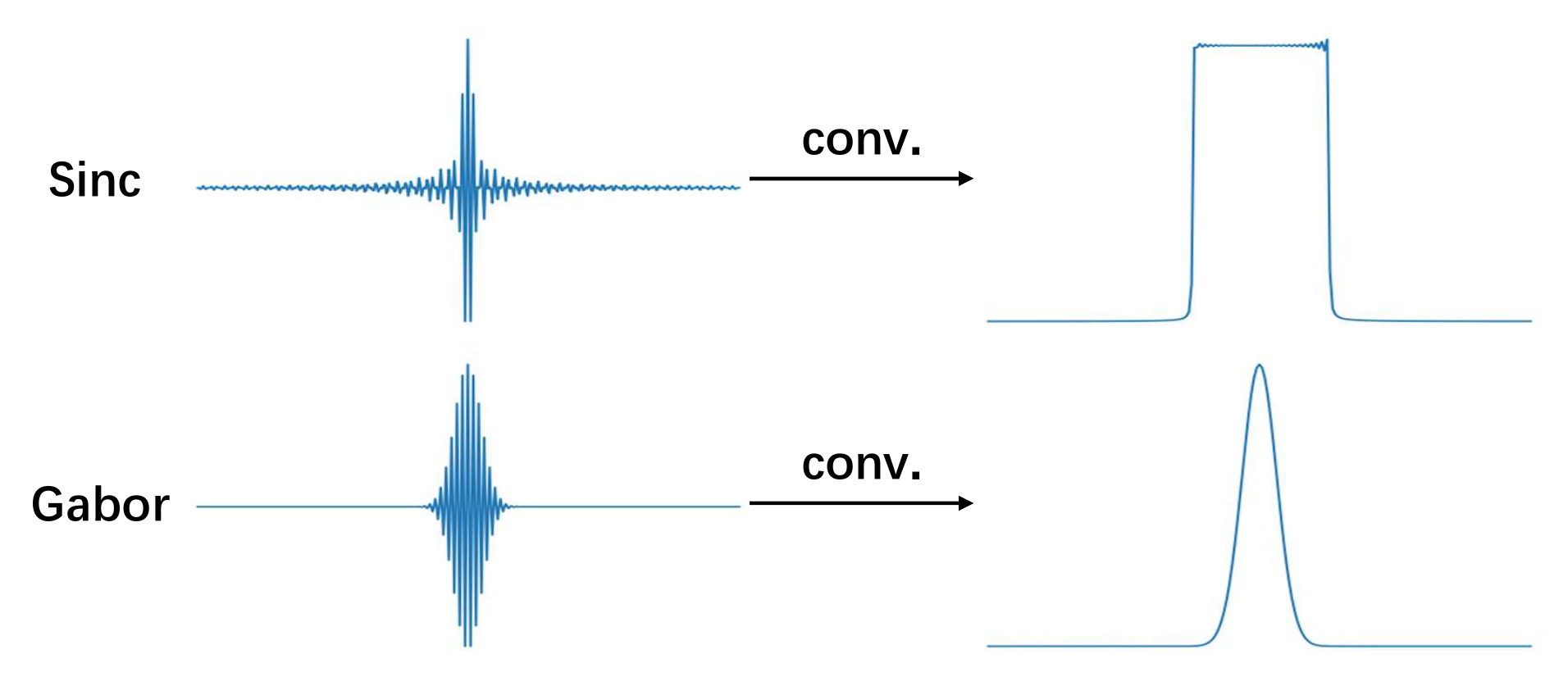}
  \caption{The convolution kernel of Sinc and Gabor filters.}
  \label{fig:pipeline}
  \vspace{-4mm}
\end{figure}

In \cite{zeghidour2021leaf}, it was shown that, without constraint on the filter weights, the frequency responses of the learned filters may exhibit spiky shapes and spread over a wide range, even to the negative frequency region, which do not appeal to our intuition. In view of these undesirable outcomes, Gabor filters were proposed to parameterize the convolution weights in \cite{zeghidour2021leaf}. In another concurrent work \cite{ravanelli2018speaker}, the Sinc-function filter was adopted for the parameterization.
The frequency responses of both types of filters approximate band-pass filters with rectangle and bell-shape respectively as shown in Fig~\ref{fig:pipeline}. In addition, the number of learning parameters is reduced to two for each filter, i.e., the center frequency and bandwidth. 


However, Sinc and Gabor filters suffer from a severe problem in convolution. As shown in Eq.~\ref{eq:FT}, the linear scaling in time gives an inverse scaling in frequency in FT. That is, in order to achieve a wide-band frequency output, the filter in time domain $f(at)$ must be narrow. Thus the filter gains away from the filter center are close to zero (see Fig~\ref{fig:pipeline}). As a consequence, a large part of the speech samples in the analysis frame are dismissed when a large convolution stride in time is used. To alleviate this problem, the stride is set to a relatively small value ($1$ in their work), which leads to the increase of the computation cost.

\begin{equation}
  f(at) \stackrel{FT}{\Longrightarrow} \frac{1}{|a|}F(\frac{\omega}{a}), a \in \mathbb{R^+}
  \label{eq:FT}
\end{equation}

\subsubsection{Filters on frequency}
\label{sssec:LFFreq}
Data-driven harmonic filters were proposed in \cite{won2020data}, where learnable filters, instead of the commonly used Mel-scale filters, were applied to the STFT output. $H$ triangular filters and $F$ harmonics of each filter are learned in order to produce a 3-dimensional feature with shape $F\times H\times T$, representing Harmonic$\times$Frequency$\times$Time as in the harmonic constant-Q transform (HCQT). State-of-the-art results across various tasks, like automatic music tagging and keyword spotting, were achieved with this trainable front-end. The learnable filters proposed in this paper can be viewed as a simplified version of these harmonic filters by omitting the harmonic term. We choose to omit the harmonic term because it is originally designed to emphasize content that has a harmonic structure, which is of more importance for tasks like music information retrieval than speaker verification. We also carried out a detailed analysis of what factors affect the degree of improvement for speaker verification, by visualizing the learned filter parameters and comparing them with those of Mel-scale filters. We empirically show that for speaker verification, the main benefit of learnable filters comes from the adjustable bandwidths, and that the learned frequency centers are similar to those in Mel-scales.

\section{Model architecture}
\label{sec:method}

\begin{figure}[t]
  \centering
  \includegraphics[width=0.9\linewidth]{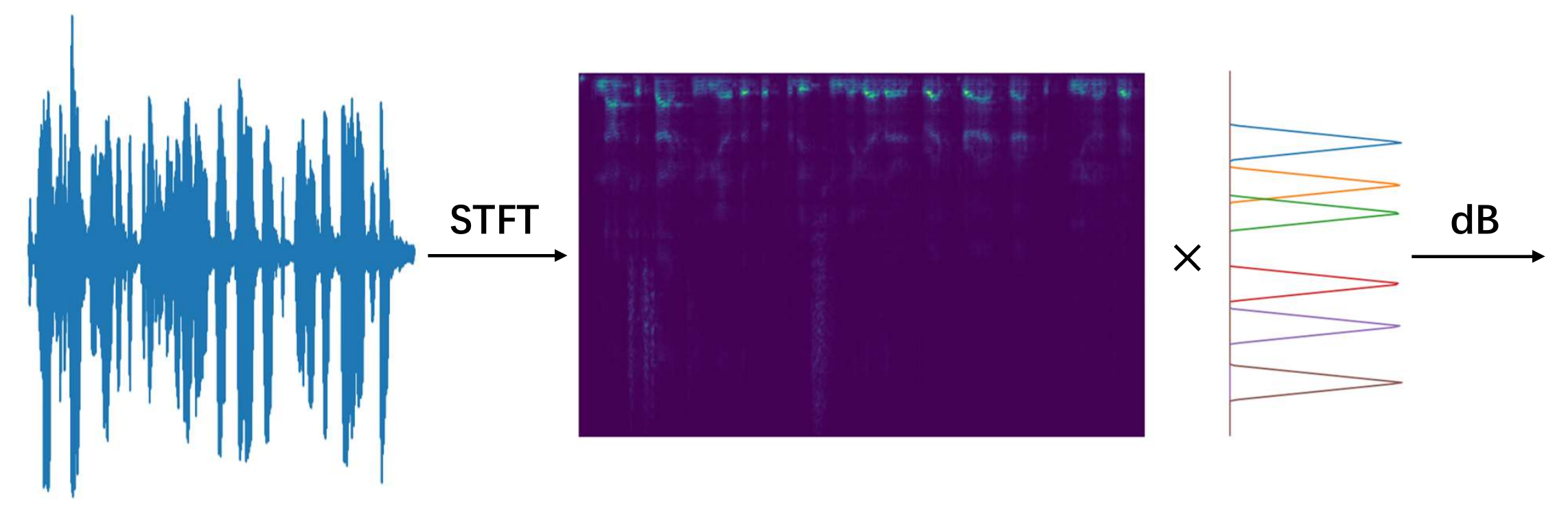}
  \caption{The learnable filters are applied on the STFT spectrum and transferred into dB scale.}
  \label{fig:LFF}
  \vspace{-4mm}
\end{figure}

\subsection{Learnable filters}
\label{ssec:LWin}
A group of learnable filter functions are applied to the STFT spectrum to generate the filter-bank output feature. 
Two types of filter functions are attempted in this work:

\noindent\emph{Triangle(T)-type:}
\begin{equation}
  \vspace{-2mm}
  w_i[n] = ReLU\{1-2\times \frac{|n-\alpha_i|}{\beta_i}\}, n \in [1, N], i \in M
  \label{eq:t_bank}
    \vspace{-1mm}
\end{equation}
\emph{Bell(B)-type:}
\begin{equation}
  \vspace{-2mm}
  w_i[n] = e^{-\frac{(n-\alpha_i)^2}{2\beta_i^2}}, n \in [1, N], i \in M
  \label{eq:b_bank}
    \vspace{-1mm}
\end{equation}
where $N$ is the number of frequency bins, $M$ is the number of filters in the learnable filter-bank. The filter's center frequency and bandwidth are determined by two learnable parameters $\alpha$ and $\beta$. The triangle-shape filter (\emph{T-type}) is defined as in Eq.~\ref{eq:t_bank}, and Eq.~\ref{eq:b_bank} defines a bell-shape filter (\emph{B-type}). Stacking of $w_i$ creates a transformation matrix $W$ of size $N \times M$. Given a spectrum with size $T \times N$ (time-frequency representation), the filter bank output is a $T \times M$ matrix obtained by multiplying the STFT spectrum with $W$, as illustrated in Fig~\ref{fig:LFF}. The output values are transformed into decibel (dB) scale.

Notably, if the values of $\alpha$ and $\beta$ are specified according to the Mel-scale filters and fixed, the output features would be the log Mel-scale filter-bank features (\emph{MFBANK}). While the learnable module allows flexible adjustment on the filters' locations and bandwidths to capture speaker discriminative information.



\subsection{Issues on the computation cost}
\label{ssec:cost}

For an input waveform with $l$ samples, a convolution operation with kernel $w$ and stride $s$ requires the computation cost of $O(w\frac{l}{s})$. The used of small stride used in previous work\cite{zeghidour2021leaf,ravanelli2018speaker} places a heavy computation burden. But large stride degrades their performance in the experiments. The small stride is not required in the proposed method, because the FT coefficients change little within a short time interval. The proposed method is applied on the STFT spectrum, which does not require a small stride, alleviating the convolution computation cost.




\section{Experimental setup}
\label{sec:experiment}

\subsection{Datasets}
\label{ssec:data}
Experiments are carried out on two speech datasets of different languages. The first dataset is the Voxceleb1 and Voxceleb2 \cite{nagrani2017voxceleb,chung2018voxceleb2,nagrani2020voxceleb}, in which most of the utterances are in English. The SV model is trained on the development set of Voxceleb2 (Vox.2), which contains over $1\ million$ utterances from $5,994$ speakers. Three official test sets in Voxceleb1 are used for evaluation: the cleaned original test set (Vox-O), the extended test set (Vox-E), and the hard test set (Vox-H). 

The second dataset is the CN-Celeb\cite{fan2020cn}, which consists of over $100k$ recordings in Chinese from $1,000$ speakers. We use the default train/eval split provided in the dataset. The cross-language generalization of extracted features is evaluated on CN-Celeb. All audio data are sampled at $16$ kHz and no data augmentation is applied in the experiments.


\subsection{Backbone network}
\label{ssec:backbone}

One of the main modules in an SV system is the backbone network, which takes acoustic features as input and generates the speaker embeddings. The backbone used in this work is a modified version of Time delay neural network (TDNN)\cite{snyder2018x}, which is made up of several 1D convolution layers with dilation and a statistics pooling layer. Compared with the network structure in \cite{snyder2018x}, there are three main modifications: (1) an instance normalization layer (IN)\cite{ulyanov2016instance} is added at the top, which normalizes the input features on time dimension; (2) the original pooling layer is replaced by an attentive statistics pooling layer\cite{Okabe2018}; (3) the output dimension of layer $segment7$ is set as $256$, and this layer's output is utilized as the speaker embedding. 

\subsection{Training and Evaluation}
\label{ssec:setup}

An Additive Margin Softmax Loss\cite{wang2018additive} with scale=$30$ and margin=$0.2$ is employed for speaker classification during training. The feature extraction module and the backbone network are trained jointly by an Adam optimizer\cite{DBLP:journals/corr/KingmaB14} with a batch size of 128 to minimize the classification loss. Each sample within a batch is a speech segment of 2-second long randomly cropped from an utterance. The model is trained on Vox.2 for $30$ epochs. The learning rate is initialized as $0.001$ and decayed by a ratio $0.1$ at epoch $15$, $25$ respectively.

In the evaluation, each utterance is divided into multiple 4-second duration segments, with a 3-second overlap between two neighboring segments. The average cosine similarity between the segments from the test utterance and the enrollment utterance is used as the score for verification.

\section{Results}
\label{sec:result}

\subsection{LFF vs. MFBANK}
The number of frequency bins ($N$ in Eq.~\ref{eq:ft}) in STFT is set to $512$, and the number of band-pass filters in both \emph{LFF} and \emph{MFBANK} is $64$. The Equal Error Rate (EER\%) results on the Voxceleb1 test sets are summarized in Table \ref{tab:mfbank_lff}. We can see that \emph{LFF} outperforms \emph{MFBANK} on all test sets, and that $T$-type and $B$-type learnable filters do not show significant differences in the model performance.

\begin{figure}[t]
\centering
\begin{subfigure}{.23\textwidth}
  \includegraphics[width=1\linewidth]{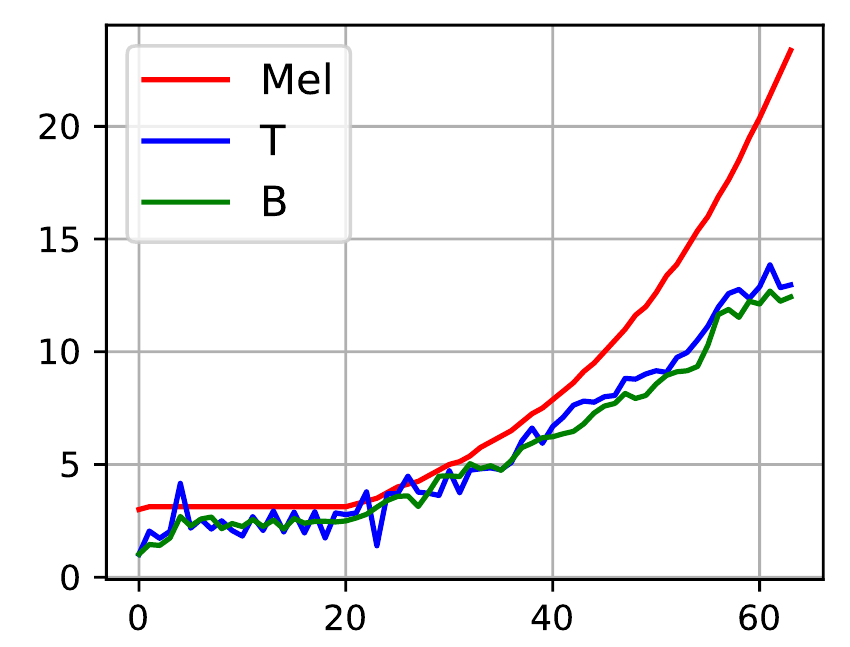}
  \caption{}
  \label{fig:sub-bw}
\end{subfigure}
\hfill
\begin{subfigure}{.23\textwidth}
  \includegraphics[width=1\linewidth]{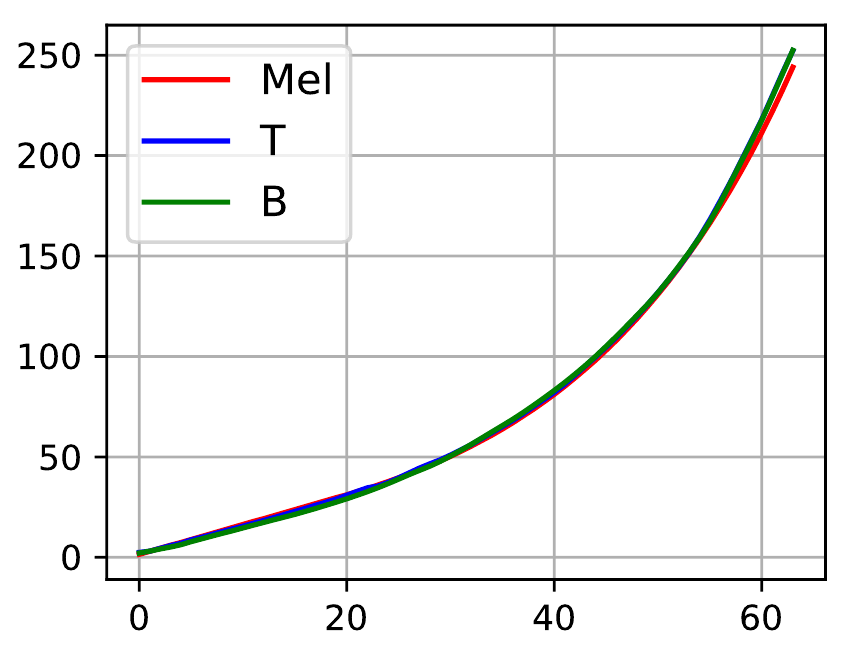}
  \caption{}
  \label{fig:sub-bc}
\end{subfigure}
\caption{The y-axis of (a) represents the filters' bandwidths, (b) represents the center frequencies of the filters.}
\label{fig:filter}
\vspace{-2mm}
\end{figure}

To understand what the filters have learned in the training, we analyzed their parameters. The bandwidths and center frequencies are plotted in Fig.~\ref{fig:filter}. The $x$-$axis$ of (a) and (b) represents the index of the filter, ranging from 0 to 63 (i.e. there are $64$ filters in total). The $y$-axis represents the learned parameter value. It can be observed that the bandwidths and center frequencies of the \emph{T-type} filters are highly close to \emph{B-type}. Compared with Mel-scale filters in Fig.~\ref{fig:sub-bw}, the learned filters have smaller bandwidths in the whole frequency region, suggesting that the narrow-band filters are more appropriate for extracting speaker-related characteristics. On the other hand, the center frequencies of learned filters are surprisingly similar to the Mel-scale filters, as shown in Fig.~\ref{fig:sub-bc}.

\begin{table}[t]
  \caption{The EER($\%$) of TDNN on Voxceleb 1.}
  \label{tab:mfbank_lff}
  \renewcommand{\arraystretch}{0.95}
  \setlength\tabcolsep{12pt}
  \centering
  \begin{tabular}{cccc}
    \toprule
    {Model} & Vox-O & Vox-E & Vox-H  \\
    \midrule
    \midrule
    {MFBANK}              & $2.26$       & $2.19$      & $3.81$        \\
    \midrule
    {LFF-T}   & $\bm{2.20}$  & $\bm{2.13}$ & $\bm{3.78}$   \\
    \midrule
   {LFF-B}   & $2.24$       & $\bm{2.13}$ & $3.80$        \\
    \bottomrule
 \end{tabular}
 \vspace{-2mm}
\end{table}

\begin{table}[t]
  \caption{The EER($\%$) of TDNN on Voxceleb 1. The second row gives the convolution stride and pooling stride}
  \label{tab:baseline_result1}
  \renewcommand{\arraystretch}{0.95}
  \centering
  \begin{tabular}{cccc ccc}
    \toprule
    \multirow{2}{*}{Model} & \multicolumn{3}{c}{Vox-E} & \multicolumn{3}{c}{Vox-H}  \\
    \cmidrule{2-4}\cmidrule{5-7}
    & $160$/$1$ & $80$/$2$ & $40$/$4$ & $160$/$1$ & $80$/$2$ & $40$/$4$  \\
    \midrule
    \midrule
    Sinc   & $3.15$ & $2.66$ & $2.50$ & $5.49$ & $4.73$ & $4.47$  \\
    \midrule
    Gabor  & $2.38$ & $2.22$ & $2.19$ & $4.20$ & $3.93$ & $3.87$  \\
    \midrule
    LFF-T  & $\bm{2.13}$ & $2.14$ & $2.16$ & 
                                              $\bm{3.78}$ & $3.81$ & $3.86$  \\
    \midrule
    LFF-B   & $\bm{2.13}$ & $\bm{2.10}$ & $\bm{2.15}$ & 
                                              $3.80$ & $\bm{3.79}$ & $\bm{3.80}$  \\
    \bottomrule
 \end{tabular}
  \vspace{-2mm}
\end{table}


\subsection{Performance v.s. computation cost}
In this section, the proposed method \emph{LFF} is compared with two other learnable front-end features, \emph{Gabor-conv} and \emph{Sinc-conv}, which take raw waveform as input. \emph{Gabor-conv} is modified from LEAF, in which the learnable pooling is disposed of and the learnable normalization is replaced by the logarithmic function. For a fair comparison, the kernel size of the first convolution layer for the latter two methods is set as $400$ with a stride of $160$, corresponding to the classic STFT configuration of window length $25ms$ and window shift $10ms$. The output feature dimension of all feature extraction models is fixed at $64$.

However, under this configuration, \emph{Gabor-conv} and \emph{Sinc-conv} are unlikely to achieve good performance because they require a small stride to cover wide-band frequency information. We then decrease the stride of the convolution layers in \emph{Gabor-conv} and \emph{Sinc-conv} by half. In the meanwhile, the window shift of STFT for \emph{LFF} is also reduced by half to allow fair comparison. A max-pooling layer is applied behind the convolution to tailor the length of the output. 

\begin{figure}[t]
\centering
\begin{subfigure}{.23\textwidth}
  \includegraphics[width=1\linewidth]{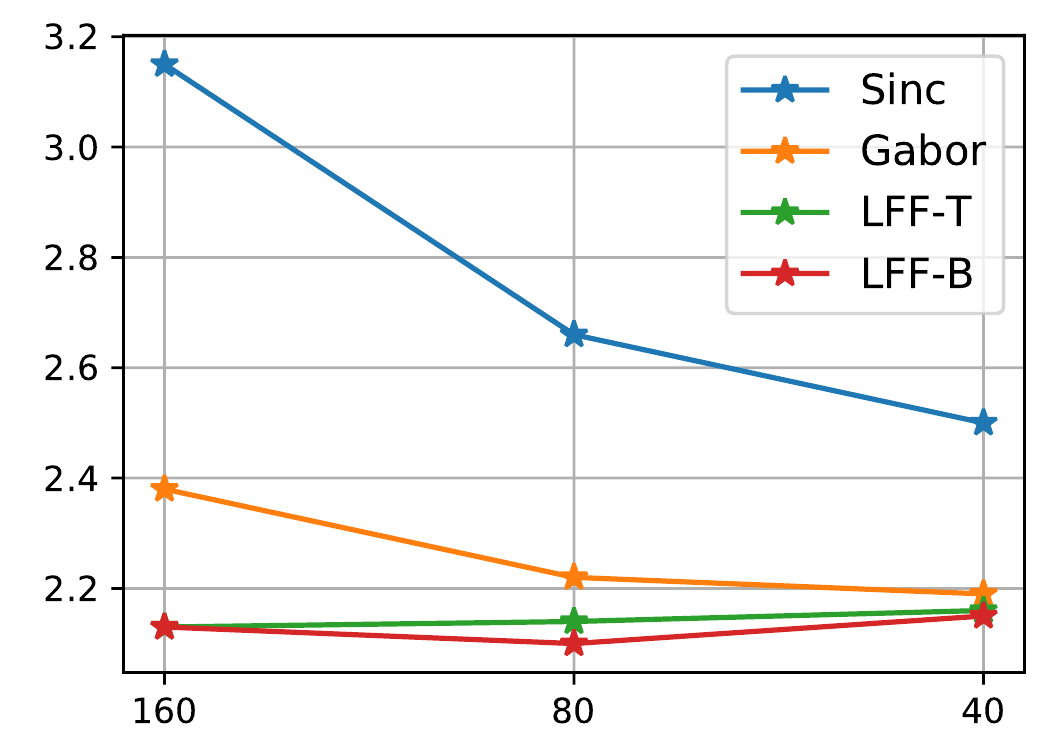}
  \caption{}
  \label{fig:sub-easy}
\end{subfigure}
\hfill
\begin{subfigure}{.23\textwidth}
  \includegraphics[width=1\linewidth]{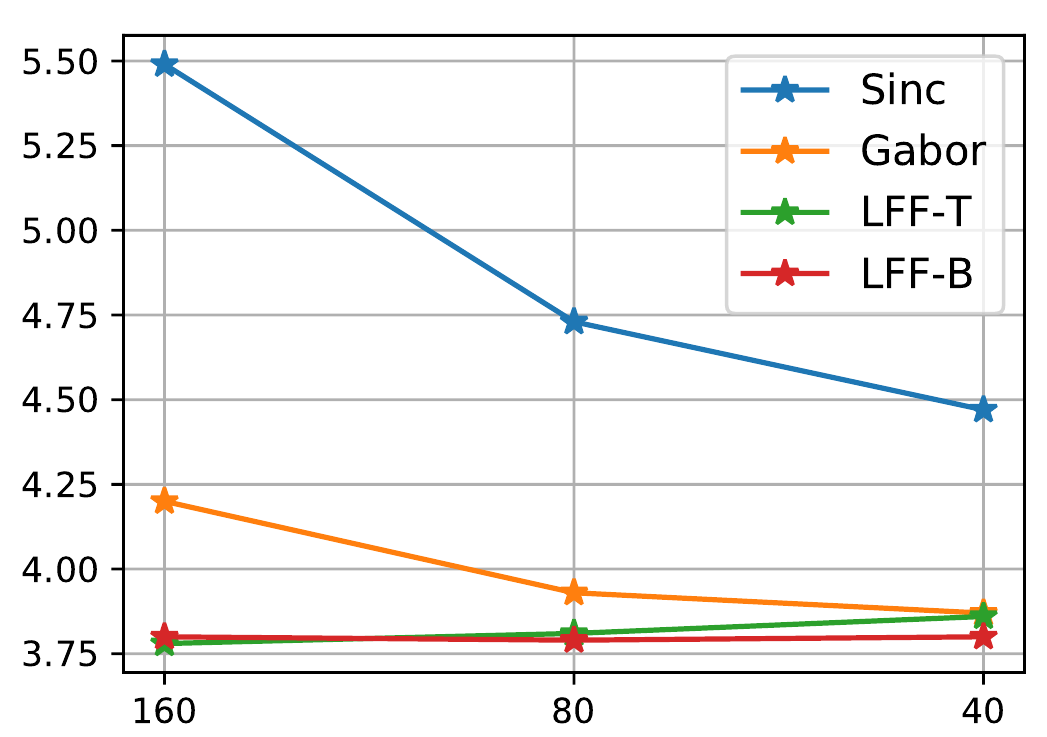}
  \caption{}
  \label{fig:sub-hard}
\end{subfigure}
\caption{The y-axis represents the EER(\%) and the x-axis represents the convolution stride. (a) Vox-E, (b) Vox-H.}
\label{fig:vox_stride}
  \vspace{-2mm}
\end{figure}

The impact of stride/window-shift on the performance of the above three learnable feature extraction models is depicted in Figure \ref{fig:vox_stride} and Table \ref{tab:baseline_result1}. Only \emph{Vox-E} and \emph{Vox-H} are shown here because they are of a much larger size than \emph{Vox-O} and therefore more representative. It is noted that both \emph{Gabor} and \emph{Sinc} depend on a small stride to obtain a good performance, but it also implies a higher computation cost. On the contrary, \emph{LFF} is not sensitive to the stride because it is directly applied on the STFT spectrum. Compared with $Gabor$ and $Sinc$, the proposed method gives superior performance under stride $160$, meaning it would be more preferable if computing resource is limited.

\begin{table}[t]
  \caption{Performance on different backbones and datasets}
  \label{tab:EER_ecapa_cn}
  \renewcommand{\arraystretch}{0.95}
  \centering
  \begin{tabular}{cclc}
    \toprule
    \textbf{Dataset}                        &   \textbf{Backbone}           &   \textbf{Feature}  & \textbf{EER(\%)}  \\
    \midrule
    \midrule
     \multirow{2}{*}{\shortstack[l]{Vox-O.}}    &   \multirow{2}{*}{\shortstack[l]{ECAPA.}}   & MFBANK             &                                                 $1.25$    \\
                                            &                                           & LFF-T              &                                                 $\bm{1.20}$    \\
    
    \midrule
    \multirow{4}{*}{\shortstack[l]{CN.}}    & \multirow{2}{*}{\shortstack[l]{TDNN}}    & MFBANK             &                                                                                         $14.59$    \\
                                            &                                           & LFF-T              &                                               $\bm{14.31}$    \\
                                            \cmidrule{2-4}
                                            & \multirow{2}{*}{\shortstack[l]{ECAPA.}}   & MFBANK             &                                                 $13.13$    \\
                                            &                                           & LFF-T              &                                                 $\bm{13.09}$    \\
    \bottomrule
  \end{tabular}
    \vspace{-2mm}
\end{table}

\begin{table}[t]
  \caption{The EER($\%$) of TDNN on Voxceleb 1.}
  \label{tab:result1}
  \renewcommand{\arraystretch}{0.95}
  \centering
  \begin{tabular}{clccc}
    \toprule
    \multicolumn{2}{c}{Model} & Vox-O & Vox-E & Vox-H  \\
    \midrule
    \midrule
    \multicolumn{2}{c}{MFBANK}              & $2.26$       & $2.19$      & $3.81$        \\
    \midrule
    \multirow{3}{*}{LFF-T}  & $\lambda=0$   & $\bm{2.20}$  & $\bm{2.13}$ & $\bm{3.78}$   \\
                            & $\lambda=0.1$ & $2.21$       & $2.15$      & $\bm{3.78}$   \\
                            & $\lambda=0.2$ & $2.22$       & $2.20$      & $3.94$        \\
    \midrule
    \multirow{3}{*}{LFF-B}  & $\lambda=0$   & $2.24$       & $\bm{2.13}$ & $3.80$        \\
                            & $\lambda=0.1$ & $2.21$       & $2.15$      &  $3.79$       \\
                            & $\lambda=0.2$ & $2.22$       & $2.17$      & $3.86$        \\
    \bottomrule
 \end{tabular}
  \vspace{-2mm}
\end{table}






\subsection{Model generalization}
\label{ssec:generalization}
We evaluate the generalization ability of the proposed \emph{LFF} by testing it on another network architecture: ECAPA-TDNN\cite{desplanques20_interspeech}, and another dataset: the CN-Celeb dataset. 

ECAPA-TDNN involves the Res2Net structure into the TDNN and gives a state-of-the-art performance. The $512$-channel ECAPA-TDNN is used in the experiments and the training process is similar to TDNN as described in Section~\ref{ssec:setup}. For the CN-Celeb dataset, the size of which is much smaller than Voxceleb, the speaker embedding dimension is decreased to $128$, and a dropout layer with $p=0.3$ is added before the embedding layer to alleviate overfitting. The number of training epochs is also reduced to $15$.

Table.~\ref{tab:EER_ecapa_cn} compares the performance of \emph{LFF} and \emph{MFBANK} on the previously described two backbone architectures and datasets. It shows that features from \emph{LFF-T} give consistently better results than \emph{MFBANK}, suggesting that the proposed learnable frequency filters generalize well on different network architectures and languages.

\subsection{CNN filter-banks}
\label{ssec:LCnn}

The proposed approach aims to extract useful features from the STFT spectrum. To evaluate whether the information extracted by CNN from raw waveform can complement the features of learned filters and improve the performance, a one-layer CNN is applied. The input waveform is normalized by mean and standard deviation first (achieved by an IN layer). The normalized waveform is processed by the convolution layer to generate $T \times \lambda M$ output features, where $T$ denotes the output time length. $\lambda$ is a hyperparameter smaller than $1$ and controls the relative contributions from the CNN and \emph{LFF}. For the output feature with size $T \times M$, $(1 - \lambda) M$ channels of it are generated from \emph{LFF}. The convolution kernel size and stride for both CNN and \emph{LFF} are set as $400$ and $160$ respectively.

The results are shown in Table~\ref{tab:result1}. It can be observed that involving CNN in the feature extraction does not improve the performance. It indicates that vanilla convolution can not provide additional information for SV within a low computation cost (large stride and small output dimension). The design of CNN for feature extraction from raw waveform requires further careful investigation.

\section{Conclusions}

A learnable feature extraction front-end for SV, named \emph{LFF}, has been developed and evaluated. The model consists of a group of filters with learnable bandwidth and center frequency, and the filters are applied on the STFT spectrum to extract filter-bank features. Two different filter shapes are investigated in the experiments and they give similar performances in SV. Compared with conventional Mel-scale filters, the learned filters exhibit narrower bandwidths. The proposed method can be implemented with low computation cost and performs better than two other learnable features \emph{Gabor}, \emph{Sinc} under a fair comparison in the experiments.

\section{Acknowledgements}

Jingyu LI is supported by the Hong Kong Ph.D. Fellowship Scheme of the Hong Kong Research Grants Council.

\bibliographystyle{IEEEtran}

\bibliography{mybib}


\end{document}